\documentstyle[sprocl,epsfig]{article}

\newcommand{\bce}{\begin{center}}
\newcommand{\ece}{\end{center}}
\newcommand{\beq}{\begin{equation}}
\newcommand{\eeq}{\end{equation}}
\newcommand{\ket}[1]{| {#1} \rangle}

\def\lsim{\mathrel{\rlap{\lower4pt\hbox{\hskip1pt$\sim$}}
    \raise1pt\hbox{$<$}}}         
\def\gsim{\mathrel{\rlap{\lower4pt\hbox{\hskip1pt$\sim$}}
    \raise1pt\hbox{$>$}}}         

\def\be{\begin{equation}}
\def\ee{\end{equation}}
\def\bea{\begin{eqnarray}}
\def\eea{\end{eqnarray}}

\begin{document}

\title{NONPERTURBATIVE EFFECTS IN THE PROTON SEA}

\author{W. SCH\"AFER, J.SPETH}

\address{Institut f\"ur Kernphysik, Forschungszentrum J\"ulich 
D-52425 J\"ulich, Germany\\E-mail: Wo.Schaefer@fz-juelich.de} 


\maketitle\abstracts{We revisit the evaluation of the pionic mechanism of the 
$\bar{u}-\bar{d}$-asymmetry in the proton structure function.
Our analysis is based on the unitarity relation 
between contributions of different mechanisms
to the inclusive particle production and the total
photoabsorption cross-section (i.e. the proton structure function).
We reanalyze the role of isovector reggeons in inclusive production of
nucleons and Delta isobars in hadronic reactions. A rather
large contribution of reggeon-exchange induced production of Delta isobars
is found. This leaves much less room for the pion-exchange induced mechanism of
$\Delta$ production and provides a constraint on the $\pi N \Delta$ form factor.
The production of leading pions in proton-proton collisions 
puts additional constraints on the $\pi NN$ vertex form factors.
All these constraints are used then to estimate the pion content
of the nucleon and allow to calculate parameter-free the $x$-dependence
of $\bar d - \bar u$.
We discuss the violation of the Gottfried Sum Rule and $\bar d$-$\bar u$
asymmetry and compare to the one obtained from the E866 experiment
at Fermilab.}

\section{Introduction}

Since the discovery of the Gottfried sum rule
violation \cite{A91} there has been a long ongoing
discussion on the $\bar{d}-\bar{u}$
asymmetry in the nucleon sea. Considerable atttraction
has been received recently by the E866 experiment at
Fermilab\cite{E866}, which provided the first detailed
measurement of the Bjorken--$x$ dependence of the 
$\bar{d}-\bar{u}$--asymmetry from the comparison
of $pp$ and $pd$ Drell--Yan production. Their striking
finding is that the asymmetry tends to vanish at
$x \sim 0.3$.

The strong observed asymmetry has no explanation
in terms of the purely pQCD dynamics, and thus
it gives important hints on the impact 
of the nonperturbative proton structure on 
the generation of the nucleons's 
sea quark content.
Following the early work of Sullivan \cite{Sul72},
a natural dynamical explanation emerges within
the framework  of the isovector meson cloud
model of the nucleon (for a recent review see
\cite{ST98}). Here a special role is played
by the pion, and it is intuitively
appealing to account for the nonperturbative 
meson/baryon structure of the proton, by including
the pion as the nonperturbative parton in
the light cone wave function of the interacting nucleon:
\beq
\ket{N}_{phys} = \ket{N}_{bare} + \ket{N \pi} + \ket{\Delta \pi} + ... \, .
\eeq
Here the pion, that is contained in the $N \pi,\Delta \pi$--Fock
states can emerge as the target in the deep inelastic
$\gamma^* N$ photoabsorption process, while the spectator baryonic
constituent $N,\Delta $ appears in the final state,
separated from the remnants of the $\gamma^* \pi$--interaction
by a rapidity gap and carrying a large fraction $z$ of the
incoming proton's lightcone momentum. 
Hence, inclusive production of baryons,
like $\gamma^* p \to X n$ is naturally described in terms
of the pionic 'partons' in the nucleon. But precisely
{\it{the same}} dynamics is supposed to
be at work, if we swap the virtual photon against
the proton projectile. This opens the possibility to
use a large body of experimental knowledge on
inclusive particle production in hadronic reactions to
constrain the meson/baryon dynamics relevant for deep
inelastic scattering.
In the following we shall demonstrate, that if
one accounts consistently for the so--derived constraints,
a satisfactory description of the Fermilab data emerges \cite{NNNSSS99}.
\begin{figure}[h]
\epsfxsize=1.\hsize
\epsfbox{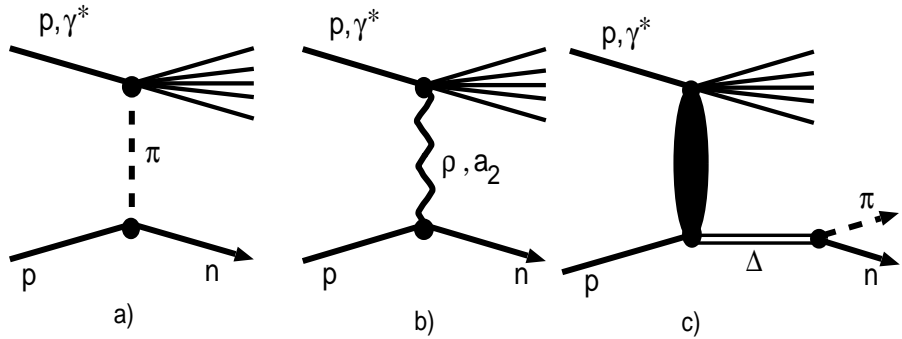}
\caption{The mechanisms for inclusive production of forward neutrons.
a) pion exchange, b) background from (reggeized) $\rho,a_2$ exchanges,
c) neutrons originating from the decay of an intermediate
$\Delta$--resonance.}
\label{fig1}
\end{figure}
\section{Inclusive production of baryons and mesons in hadronic
reactions}

\subsection{Production of forward neutrons and $\Delta$--isobars}

Following the above described strategy, we first
turn to the description of the forward neutron production
in $pp$ collisions. We take into account three
production mechanisms: the dominant pion exchange, 
the background contribution from the 
isovector exchanges $\rho,a_2$, and, finally, the
production of neutrons from the decay of an intermediate
$\Delta$--resonance (see fig \ref{fig1}).
In addition, we have to account for the distortion
of the incoming proton waves, employing standard 
methods of the generalized eikonal approximation 
\cite{NSZ97}. Following the reasoning
in \cite{NSZ97}, we apply the absorptive corrections 
only in the $pp$ scattering, they can be neglected
in the $\gamma^* p$--case.
It is important to notice,that 
the phase space of the forward neutrons includes
the kinematical boundary $z \sim 1$, which corresponds
to a large Regge parameter $s/M_X^2 \gg 1$, where
$M_X^2$ is the invariant mass squared of the inclusive
system $X$. Hence the proper formalization of the 
$\rho$--exchange mechanism should be a Regge treatment
\cite{SNS98}.

In the Regge formulation, the contribution 
to the inclusive cross section from 
the pion exchange mechanism has the form:
\beq
z{d\sigma (p \to n) \over dz dp_\perp^2} =
{g^2_{\pi pn} \over 16 \pi^2} {(-t) \over [t-m_\pi^2]^2} \cdot
F^2_{\pi NN} (t) (1-z)^{1-2\alpha_\pi (t)} \cdot \sigma^{\pi p}_{tot}(M_X^2) \, ,
\eeq
with $-t = p_\perp^2 /z +(1-z)^2m_p^2/z $, $z$ is the neutrons
Feynman--variable, which for large $z$ is equal to its lightcone
momentum fraction, $\vec{p}_\perp$ is the neutron's transverse 
momentum. Due to the proximity of the pion pole, the departure
of $\alpha_\pi (t) = \alpha' (t-m_\pi^2) \, , \alpha' \approx 0.7 \, GeV^{-2}$
 from $J_\pi = 0$ is small, 
and we can jusify taking the physical pion--nucleon cross
section $ \sigma^{\pi p}_{tot}$. 
The finite extension of the involved particles and/or the
off shell effects are accounted for by the 
form factor $F^2_{\pi NN} (t)$, in the following
we shall stick to the parametrization:
$F^2_{\pi NN} (t) = \exp[ (R^2(t-m_\pi^2))^2 ] \, , R^2 = 1.5 GeV^{-2}$, 
for different choices of the functional form, 
see the discussion in \cite{NNNSSS99}.

In fig.2 we show our results for the invariant cross section 
against the experimental data. The data shown here are at relatively 
low $p_{LAB} \approx 24 \, GeV/c$, hence the visible 
trace of the $\pi N$--resonance--region at large $z$. The $p_\perp$--
dependence nicely illustrates the interplay of the different mechanisms.
The pion exchange peaks in the region $0.7 \lsim z \lsim 0.9$,
and dominates for $p_\perp \lsim 0.3 \,  GeV$. The relative
importance of the $\rho,a_2$ exchanges grows with $p_\perp$, which
testifies to the dominant $\propto p_\perp^2$ spin--flip
component of the $\rho N$--coupling. The background contribution
from the two--step process $p \to \Delta \to n$ turns out to
be substantial only for $z \lsim 0.8$.

We took care that the background from the $\Delta$--decay is consistent
with the experimental data on forward-$\Delta$--production, to which we
turn now. Unfortunately, experimental data are rather scarce, and
often plagued by large uncertainties due to the nonresonant $\pi N$
subtraction.
We do not have any data differential
in $p_\perp$ at hand, and obviously the shape of the 
$p_\perp$--integrated $\Delta$--spectrum could be fitted by both 
extreme choices: the pure $\pi$--exchange as well as the pure $\rho$--exchange.
Here valuable information comes from the two--body reactions at high energies.
The cleanest manifestation of $\rho$--exchange is clearly the $\pi N$--charge
exchange reaction $\pi^- p \to \pi^0 n$ and $\pi^+ p \to \pi^0 \Delta^{++}$,
for which experimental data are available \cite{IW77}.
Assuming Regge factorization, we can now relate the $\rho$--exchange contribution
in the $p \to n$ transition from our earlier considerations to
the  $\rho$--exchange in $p \to \Delta$.
The result is a surprisingly large $\rho$--exchange contribution,
which does not leave much room for the pionic contribution (see fig 3.).
\begin{figure}[h]
\epsfxsize=0.7\hsize
\epsfbox{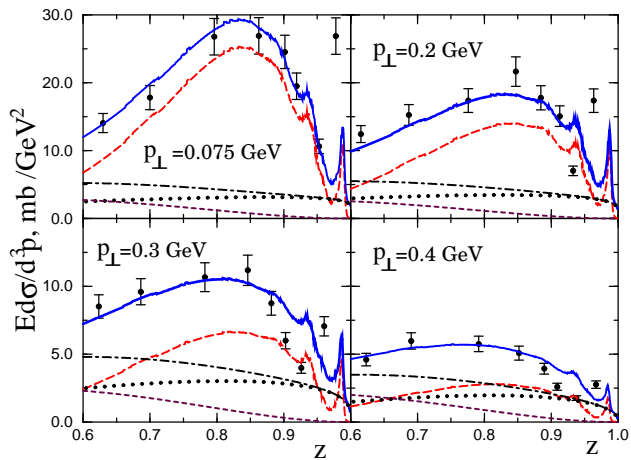}
\caption{ Invariant cross section for the reaction
$pp \to nX$ at $p_{LAB} = 24 \, \mbox{GeV/c}$.
The experimental data are taken from \protect\cite{blobel}.
The long dashed curve shows the contribution from the pion exchange;
the dotted curve is the $\rho,a_2$-exchange
contribution, and the dashed curve shows the contribution from the two
step process $p \to \Delta \to n$. In additionally we present the sum of
the two background contributions as the dot-dashed line.
Finally the solid curve represents the sum of all components.
}
\end{figure}
\begin{figure}[h]
\epsfxsize=0.6\hsize
\epsfbox{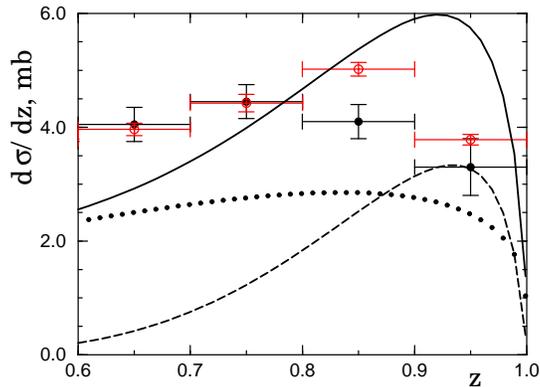}
\caption{ Differential cross section
$d\sigma / dz$ for the reaction $pp \to \Delta^{++} X$
at $p_{LAB} = 400 \, \mbox{GeV/c}$.
The dashed curve is the contribution from pion-exchange; the
dotted curve shows the $\rho ,a_2$ contribution.
Shown by the solid curve is the sum of the two.
Experimental data are taken from \protect\cite{LEBC-EHS}(open circles),
and \protect\cite{ABCDHW} (filled circles).}
\end{figure}

\subsection{Constraints from forward pion production}

The good description of the forward baryon production can make us
confident, that we have found a proper description of the 
$\pi N, \pi \Delta$--fock states in the region where the baryon
carries a large momentum fraction $z\sim 0.6 \div 1$. 
However we should be aware, that the forward baryon production
data do not put any constraint on the region $ z\lsim 0.6$;
in such a kinematical region we cannot expect the meson/Reggeon
exchanges to be the dominant reaction mechanisms. On the
other hand, in the Fock--state picture a baryon carrying
a small momentum fraction $z_B \ll 1$ is accompanied
by the meson carrying large $z_M = 1-z_B \sim 1$.
Hence, the natural place to check the consistency of the
Fock--state parameters is the production of forward, $z_\pi \sim 1$, pions.
The important message from the experimental data (fig.4) is:
there are {\it{no pions carrying large momentum fractions}} $0.6 \lsim z \lsim 1$
observed. This places a severe constraint on the cutoff in the 
employed $\pi NN$--formfactor. (Notice that we are not after 
a description of the forward pion production for $z \lsim 0.5$).
The specific formfactor choice cited above leads to the
middle solid curve in fig.4. Incidentally, the dotted line on 
fig.4 shows the result of a previously emloyed formfactor \cite{HSS96} that
also did a reasonable job on forward neutrons, but does
not respect the pion production data.

\begin{figure}[h]
\epsfxsize=0.6\hsize
\epsfbox{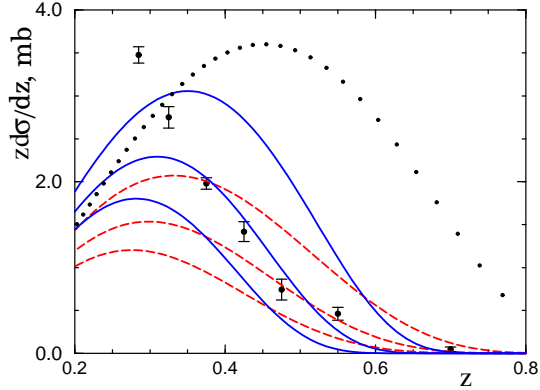}
\caption{ Differential cross section $zd\sigma /dz (pp \to \pi^0 X)$
at $p_{LAB} = 400 \, \mbox{GeV/c}$. The data are taken from \protect\cite{LEBC-EHS}.
The dotted curve shows a prediction of the model \protect\cite{HSS96}.
The solid curves were calculated with a 'Gaussian' form factor
$F_{\pi NN}(t) = \exp (-\left[R_G^2(t-m_\pi^2)\right]^2)$.
The curves in the figure correspond to $R_G^2 = 1.0\, ,1.5 , and \, 2.0 \; GeV^{-2}$
(from top to bottom). The dashed curves were calculated with an exponential
form factor $F_{\pi NN}(t) = \exp (R_E^2(t-m_\pi^2))$.
The curves in the figure are for $R_E^2 = 1.0\, ,1.5,and \, 2.0 \; GeV^{-2}$
(from top to bottom).}
\end{figure}

\section{Regge mechanisms vs. Fock--state picture}
In the introduction we stressed the relevance of the
pion as the nonperturbative parton in the
light--cone wave function of the interacting proton.
Here the question arises what kind of Fock--state should then
be associated with the $\rho,a_2$-Reggeon exchange
production mechanisms, if any? 
In other words, knowing the reaction mechanisms that
populate several inclusive channels, what can we
say about their impact on the total cross section?
We remind, that one of the firm predictions of Regge theory
is a very specific phase of the amplitudes. 
While the inclusive cross section is calculated from 
the modulus of the amplitude squared, $|A|^2$, it is the product
of two amplitudes, $A \cdot A$, that enters the evaluation
of the total cross section. The result may be
summarized by generalized AGK cutting rules \cite{NNNSSS99,AGK74}:
for the purely real pion exchange amplitude, this implies
that inelastic interactions of the projectile 
with the pions in the target hadrons enhances the 
total cross section, whereas the $\rho,a_2$--exchanges,
which have a phase $\sim (1+i)$, happen to give
a vanishing contribution to the total cross section.
Hence, it does not make any sense, to identify
the Reggeon exchange mechanism in inlusive reactions in terms
of a meson/baryon--Fock state of the proton.
We want to point out that there is no mystery connected with
such a different impact of the phases on different observables.
For instance in the inclusivex $pp \to pX$, diffractive 
channels (imaginary amplitude) 
dominate for $z \sim 1$, and it is well known
that the opening of diffractive channels comes along 
with the {\it{absorptive correction}} to the total
cross section, which is {\it{negative}} \cite{AGK74}.

\section{The $\bar{d} - \bar{u}$ asymmetry from pions in the nucleon}

Following our reasoning above, we include the $\pi N$, $\pi \Delta$--
Fock states only. They lead to a contribution to 
the total $\gamma^* p$--cross section -- i.e. the quark/
antiquark--distributions in the proton given by
\beq
\Delta^{\pi} q (x,Q^2) = \int_x^1 {dz_\pi \over z_\pi} 
f_\pi (z_\pi) \cdot q^\pi (x/z_\pi, Q^2) \, ,
\eeq
with the flux of pions which is determined through
\beq
 f_{\pi} (z_\pi) = \int d^2 \vec{p}_\perp E{d\sigma (p \to n) 
\over \sigma_{tot}^{\pi p} d^3\vec{p}}  \, ,
\eeq
and similarly for the $\pi \Delta$--Fock state.
Besides the Fock--state parameters determined in our analysis above,
we need as an input the quark distributions in the pion.
Notice, that for the calculation of $\bar{d}(x)-\bar{u}(x)$ only
the pion's valence distributions enter. These are reasonably well
constrained down to $x \sim 0.2$ from the Drell--Yan experiments,
for definiteness we take the GRV--parametrization \cite{GRV92} .
We obtain a total multiplicity of pions in the proton associated
with the $\pi N$--state of $n_{\pi N} \sim 0.21 \div 0.28$ and
of $n_{\pi \Delta} \sim 0.03 \ll n_{\pi N}$ for the $\pi \Delta$--contribution.
This translates to a Gottfried sum of 
$0.21 \lsim S_G \lsim 0.25$, compared to the NMC determination\cite{A91} $S_G =
0.235 \pm 0.026$
Our result for $\bar{d} - \bar{u}$ is shown in fig 5, we observe
that the asymmetry is essentially driven by the $\pi N$--state.
Notice that our good agreement in the region $x \gsim 0.2$ owes
to the fact that we have no contributions from hard $z_\pi \gsim 0.5$
pions in the nucleon.

\section{Summary}

Our reanalysis of the pionic contribution to the $\bar{d} - \bar{u}$--
asymmetry has been based on a unified treatment of the inclusive production
of leading nucleons, $\Delta$'s and pions in hadronic high energy collisions.
We paid special attention to the background contributions
in the leading baryon production, and clarified the relation of
Reggeon--exchange mechanisms and the Fock--state picture of
the interacting nucleon. Severe constraints on our parameters were
found to arise from the forward pion data. A good agreement 
with the recent E866 data on  $\bar{d} - \bar{u}$ was found.

Finally we want to point at another spin--off of our analysis:
namely one may use the pions in the nucleon as 
targets in deep inelastic scattering and hence obtain
information on its structure function at small $x$, unexplored
by the Drell-Yan experiments\cite{HLNSS94}. Our findings for 
the relevant background
contributions show that this task is well feasible.

\begin{figure}[h]
\epsfxsize=0.7\hsize
\epsfbox{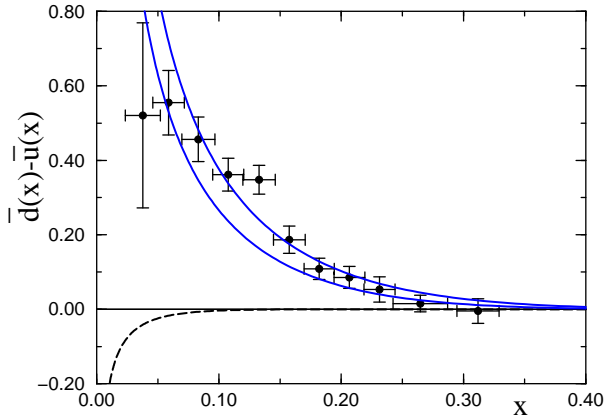}
\caption{Flavour asymmetry $\bar{d} (x) - \bar{u} (x)$ at $Q^2 = 54 \, \mbox{GeV}^2$. Experimental data are from E866 \protect\cite{E866}. The solid curves show the contribution from the $\pi N$-Fock state and were calculated for Gaussian form-factors; the upper curve belongs
to $R_G^2 = 1 \, \mbox{GeV}^{-2}$, the lower one to  $R_G^2 = 1.5 \, \mbox{GeV}^{-2}$
The dashed line shows the contribution of the $\pi \Delta$-Fock state.
}
\end{figure}

\section*{Acknowledgments}
We wish to thank C. Carlson and A. Radyushkin for invitation.
W.S. is indebted to INT/JLab for the financial support.

\end{document}